\documentclass[prl,twocolumn,showpacs,preprintnumbers,amsmath,amssymb]{revtex4}
\usepackage{bm}
\usepackage{graphicx}
\usepackage{bm}
\begin{document}
\preprint{aps-123}
\title{Observation of $3$D Heisenberg$-$like ferromagnetism in single crystal La${_{0.875}}$Sr${_{0.125}}$MnO${_3}$}
\author{Sunil Nair} 
\author{A. Banerjee}
\altaffiliation{Corresponding Author}
\email{alok@iucindore.ernet.in}
\author{A. V. Narlikar}
\affiliation{Inter University Consortium for D.A.E. Faclilites\\ University Campus, Khandwa Road. Indore. India 452 017.}
\author{D. Prabhakaran and A. T. Boothroyd}
\affiliation{Department of Physics, Oxford University\\ Oxford, OX1 3PU, United Kingdom}
\date{\today}
\begin{abstract}
We report measurements and analysis of the magnetic critical phenomena in a single crystal of La${_{0.875}}$Sr${_{0.125}}$MnO${_3}$. The critical exponents associated with the ferromagnetic transition have been determined from ac susceptibility and dc magnetisation data. Techniques like the Kouvel$-$Fischer plots, modified Arrots Plots and the critical isotherm analysis were used for this purpose.  The values of the the exponents $\gamma$, $\beta$ and $\delta$ obtained are found to match very well with those predicted for the $3$D Heisenberg model.  Our results are consistent with recent numerical calculations and suggest that though the double exchange interaction is driven by the motion of conduction electrons, the effective magnetic interaction near the transition is renormalized to a short range one. 
\end{abstract}
\maketitle
A universality class (based on the dimensionality of the lattice and the order parameter) is assigned to a system depending on the values of the assymptotic critical exponents that characterises the phase transition\cite{stanley}.  This practise of assigning such universal classes based on theoretical spin$-$spin interaction models (like the Heisenberg, mean field or the Ising ) has been extremely useful in trying to discern the intricacies of magnetic transitions in real systems.

The nature of the para$-$ferro magnetic phase transition in mixed valent manganites of the general formula La${_{1-x}}$A${_x}$MnO${_3}$ (where A is a divalent atom) is an area where the findings of experiments are yet to converge with theoretical models. The nature of the phase transitions in this class of materials was qualitatively explained by the Double exchange (DE) interaction\cite{zener}, which originates from the motion of conduction electrons, and its coupling with spins localised at the lattice.  Thus these materials provide us with a unique class of strongly correlated electron systems where the lattice, charge and spin degrees of freedom are strongly coupled.  It is interesting to determine how the kinetics of electrons which favour extended states affect the critical fluctuations and renormalizes the DE interaction in the vicinity of the phase transition.

Though metallicity and ferromagnetism run concurrently in the basic DE model, perturbations like large lattice distortions can cause the mixed valent manganites to be insulating.  This was aptly illustrated in the optimally doped system La${_{0.67}}$Mg${_{0.33}}$MnO${_3}$, where no metallicity was observed in the vicinity of the ferromagnetic ordering temperature\cite{will1}.  Though the values of the exponents determined were seen to match very well with the isotropic $3$D Heisenberg model, lack of metallicity in this system implies that this would not be a suitable candidate to establish the nature of ferromagnetism in a canonical DE system.

As far as \textit{metallic} DE systems are concerned, the results of critical exponent measurements have been extremely inconclusive.  In  systems with hole doping $> 10\% $, which is necessary to observe a DE driven phase transition, the values of the critical exponent associated with magnetisation ($\beta$) is seen to vary from $0.37$\cite{ghosh} to $0.50$\cite{mohan}.  This encompasses the values of the exponent as predicted by the $3$D Heisenberg , the $3$D Ising and the mean field models, making it difficult to ascertain the universality class to which these systems belong.  Besides the above mentioned values which were determined using bulk magnetic measurements, the values of the critical exponents determined using other techniques like microwave absorption\cite{lofland}, microwave surface impedance\cite{schwartz} and neutron diffraction\cite{martin} have also varied, adding to the ambiguity in this issue.

The quality of the samples used in the study of criticality in this class of materials is a probable culprit for the present state of affairs.  These samples are notoriously inhomegenous, thus causing a smeared  transition (with a distribution of T${_C}$'s), which leads to erroneous determination of the assymptotic critical exponents.  Thus, it is imperative that homogenous single crystals be used for experimental investigations of this nature.

In an attempt to resolve the inconsistencies existing in literature, and to determine the nature of ferromagnetism in  metallic DE manganites, we have done ac susceptibility and dc magnetisation measurements in the critical region of a high purity single crystal of orthorhombic La${_{0.875}}$Sr${_{0.125}}$MnO${_3}$. The samples were prepared by the standard Solid State Route using $99.999\%$ pure La${_2}$O${_3}$, SrCO${_3}$ and MnO${_2}$. Single crystals were grown using a four mirror floating zone furnace under $6-8$ atmospheres of an Ar/O${_2}$ mixture.  The crystallographic phase purity was checked repeatedly by X ray diffraction, and the samples were found to be of single phase.  The oxygen stochiometry of this sample was found to be $3.00\pm 0.02$.  The chemical homogenity of the sample was checked using EPMA and variation was found to be within $1\% $. Futher details of sample preparation can be found in ref\cite{prabha}.

Dc magnetisation and ac susceptibility measurements were carried out using home made setups\cite{krsna,ashna}. In the critical region, the temperature was controlled to an accuracy of $0.01$K using commercial Lakeshore controllers DRC$93$CA and $340$. All the measurements were made with the applied field parallel to the ab plane of the crystal.  The low field linear region of M$-$H scans were used to determine the value of the demagnetisation factor, which in our case was found to be $0.22$. All the data used for determining the critical exponents were duly corrected for the demagnetisation fields.

The composition La${_{0.875}}$Sr${_{0.125}}$MnO${_3}$ has been extensively studied in the past\cite{gross}, and is reported to show two transitions in reasonably close succession, a paramagnetic insulator$-$ferromagnetic metal transition, followed by another low temperature transition to an insulating state. This state is now being explained by invoking polaronic\cite{yamada} and orbital ordering\cite{shen} models.  However, in this study, we have concentrated on the higher temperature transition which is likely to be driven by the DE interaction.  Figure 1 shows the real part of the first order ac susceptibility and the dc magnetisation measured in this sample where the two transitions mentioned above are clearly seen.

Conventional methods of analysing the critical region involves the use of Arrot plots \cite{arrot}.  These M${^2}$ Vs H/M isotherms should be straight lines in the critical region, with a zero intercept at T$=$T${_C}$.  The intercepts for T$<$T${_C}$ and T$>$T${_C}$ is then used to determine the saturation magnetisation (M${_S}$) and the inverse susceptibility ($1/\chi{_0}$) respectively. It is to be noted that this method implicitly assumes mean field values of the exponents, and a more general form of analysis (called the modified Arrot plots) involves plotting M${^{1/\beta}}$ Vs (H/M)${^{1/\gamma}}$ in the critical region\cite{kaul1}.  However, these methods do tend to introduce some systematic errors due to the presence of two ($\beta$ and $\gamma$) free parameters in the fitting procedure.

We have circumvented this problem by using low field ac susceptibility measurements in the critical region to determine $\gamma$.  Low field ac susceptibility is an extremely useful tool in the study of a para$-$ferro phase transition, as it enables us to directly determine the true initial susceptibility ($\chi{_0}$), which otherwise has to be estimated from the extrapolation of data measured at high applied fields\cite{kaul2} which besides supressing effects like inhomogeniety in the sample can mask the true critical beahviour of the system.  This is specially important in systems like the hole doped manganites where short range correlations are known to exist till temperatures as large as 2T${_C}$\cite{teresa}. Also, the Kouvel$-$Fischer(KF)\cite{kf} analysis can then be used to independently estimate both the transition temperature (T${_C}$) and $\gamma$.  Ac susceptibility data analysed using the KF formalism ($1/\chi{_0}d/dt(\chi{_0}{^{-1}}$) Vs T) is shown in figure $2$.  We have obtained $\gamma=1.38$ which matches very well the isotropic $3$D Heisenberg value of $1.386$.  \textit{It is to be noted that to the best of our knowledge this is the first report of a KF analysis of low field ac susceptibility in this class of materials}.  The fact that a reasonable fit could be obtained to the low field data adds credence to our claim that the sample used is homogenous and of good quality.  

Modified Arrot plots based on the equation of state
\[
(\frac{H}{M}){^{1/\gamma}} = a\frac{(T-T{_C})}{T} + bM{^{1/\beta}}
\]
is shown in figure $3$.  The value of $\gamma$ determined using the KF analysis of ac susceptibility is used, and $\beta$ is varied continously, so as to obtain isotherms almost parallel to each other in the critical region.  These isotherms are curved at low fields, as they represent averaging over domains which are magnetised over different directions\cite{aha}. However, at higher fields, the isotherms are straight lines, and the best fit was obtained with a value $\beta = 0.37$.  The value of (M${_S}$) determined by the intercepts of the  modified Arrot plots in the region T$< $T${_C}$ is then analysed using the Kouvel Fischer formalism(plotting $M{_S}(dM{_S}/dT){^{-1}}$ Vs T) to reconfirm the value of $\beta$. As is shown in fig 4, a good fit was obtained , with a value $\beta = 0.372$. 
\newline
\begin{table*}
\caption{\label{tab:table 1}Values of the exponents $\beta$, $\gamma$ and $\delta$  as determined from the Kouvel-Fischer analysis of the first order susceptibility, modified Arrot plots, and the critical isotherm.  The values determined by the earlier workers in similar metallic double exchange systems and that expected for various universal models is given for the sake of comparison. }
\begin{ruledtabular}
\begin{tabular}{cccccc}
Composition &Ref. &Technique &$\beta$ &$\gamma$ &$\delta$\\
\hline
La${_{0.875}}$Sr${_{0.125}}$MnO${_3}$$\footnotemark[1]$ &This work &Bulk Magnetisation &$0.37\pm0.02$ &$1.38\pm0.03$ &$4.72\pm0.04$\\
\hline
La${_{0.7}}$Sr${_{0.3}}$MnO${_3}$$\footnotemark[1]$ &\cite{ghosh} &Bulk Magnetisation &$0.37\pm 0.04$ &$1.22\pm 0.03$ &$4.25\pm 0.2$\\
\hline
La${_{0.7}}$Sr${_{0.3}}$MnO${_3}$$\footnotemark[1]$ &\cite{lofland} &Microwave absorption &$0.45\pm 0.05$ &$-$ &$-$\\
\hline
La${_{0.7}}$Sr${_{0.3}}$MnO${_3}$$\footnotemark[1]$ &\cite{martin} &Neutron diffraction &$0.295\pm 0.004$ &$-$ &$-$\\
\hline
La${_{0.8}}$Sr${_{0.2}}$MnO${_3}$$\footnotemark[2]$ &\cite{mohan} &Bulk Magnetisation &$0.50\pm 0.02$ &$1.08\pm 0.03$ &$3.13\pm 0.20$\\
\hline
La${_{0.8}}$Sr${_{0.2}}$MnO${_3}$$\footnotemark[1]$ &\cite{schwartz} &Microwave surface impedance &$0.45\pm 0.05$ &$-$ &$-$\\
\hline
Mean Field Model &\cite{kaul1} &Theory &$0.5$ &$1.0$ &$3.0$\\
\hline
$3$D Ising Model &\cite{kaul1} &Theory &$0.325$ &$1.241$ &$4.82$\\
\hline
$3$D Heisenberg Model &\cite{kaul1} &Theory &$0.365$ &$1.386$ &$4.80$\\
\end{tabular}
\end{ruledtabular}
\footnotetext[1]{Single Crystals}
\footnotetext[2]{Polycrystals}
\end{table*}
Fig 5 shows Ln(M) plotted as a function of Ln(H) (using the critical isotherm) , the inverse slope of which is used for determining the value of the critical exponent $\delta$. We obtain a value of $\delta = 4.72$, which matches very well with the one($\delta =4.709$) determined by using the Widom scaling equation $\delta = 1+\gamma/\beta$. A chart comparing the values of the exponents which we have obtained with those reported in literature for metallic double exchange systems is shown in table $1$.  As is clearly seen, we have obtained values which match very well with the isotropic $3$D Heisenberg values.  

It has been shown analytically\cite{ma} that if in a $3$D ferromagnet, the effective exchange interaction J(r) decays with distance (r) at a rate faster than r${^{-5}}$, then the short range (Heisenberg like) exponents are valid.  However, if J(r) decays at a rate slower than r${^{-4.5}}$, then the classical (mean field like) exponents are valid.  According to the theory of Double exchange, the effective ferromagnetic interaction is driven by the kinetics of the electrons which favour extended states, and hence one would expect a (long range) mean field like universality class.  However, recent numerical calculations\cite{furu1,furu2,furu3} show that the exponents in the DE model are consistent with those of the isotropic short range $3$D Heisenberg model.

Our results match well with these calculations and are the first of its kind to show unambigously that the para$-$ferro phase transition in metallic manganites fall into the isotropic short range $3$D Heisenberg universality class.  The consistency of the exponent values determined using low field, and high field magnetisation data ensures that the values are intrinsic to the system and are not arising due to any artifacts in the measurement or fitting processes.The $3$D Heisenberg like values of the critical exponents thus imply that the effective interactions in a canonical DE model are relatively short range near the transition.  This is important in the context of phase separation models being used to explain various intruiging features of the hole doped manganites\cite{dag}.  This intrinsic phase separation (into hole rich and hole poor regions) could possibly be the reason for the short range nature of the magnetic interactions, and hence the $3$D Heisenberg like values of the critical exponents.  

\newpage
\begin{figure}
\caption{The real part of the first order susceptibility measured as a function of temperature for the sample  La${_{0.875}}$Sr${_{0.125}}$MnO${_3}$ at a exciting frequency of $133.33$ Hz and a driving ac field of $0.92$ Oe. The inset shows the dc magnetisation measured at a field of 3.42 Oe.}
\caption{Kouvel Fischer Plots of the real part of the first order susceptibility for the sample La${_{0.875}}$Sr${_{0.125}}$MnO${_3}$. The inverse of the slope gives the value of the susceptibility exponent $(\gamma)$, and the intercept on the temperature axis provides the transition temperature $(T{_C}$).  We have obtained $\gamma = 1.38$ and $T{_C} = 186$K. Increasing non linearity cause deviations from the stright line fit close to the transition. }
\caption{Modified Arrot plots using the equation of state$(\frac{H}{M}){^{1/\gamma}} = a\frac{(T-T{_C})}{T} + bM{^{1/\beta}}$. We have used $\gamma=1.38$, and the best fit was obtained with $\beta= 0.37$.  All the isotherms measured in the critical region has not been shown for the sake of clarity. }
\caption{Kouvel Fisher analysis of the Saturation Magnetisation (M${_S}$) plotted as a function of temperature.  The inverse of the slope gives a value $\beta = 0.372$.}
\caption{LnM plotted as a function of LnH for the isotherm measured at T$=$T${_C}$. The inverse of the slope of the linear region gives a value of $\delta=4.72$.}
\end{figure}
\end{document}